%% file: ext-abstract-pp.tex
\documentclass[copyright,creativecommons]{eptcs}
\providecommand{\event}{TERMGRAPH 2018} 

\input{packages.tex}

\title{Modeling Terms by Graphs with Structure Constraints \\ (Two Illustrations)}
\author{Clemens Grabmayer
  \institute{Gran Sasso Science Institute}
  \institute{Viale F.\ Crispi 7, 67100 L'Aquila AQ, Italy}
  \email{clemens.grabmayer@gssi.it}
  }
\def\titlerunning{Modeling Terms by Graphs with Structure Constraints (Two Illustrations)}
\def\authorrunning{C. Grabmayer}

\input{macros.tex}

\begin{document}

\hypersetup{pageanchor=false}
\maketitle

\begin{abstract}
  In the talk at the workshop my aim was to demonstrate 
  the usefulness of graph techniques for tackling problems that have
  been studied predominantly as problems on the term level: 
  increasing sharing in functional programs, 
  and addressing questions about Milner's process semantics for regular expressions. 
  For both situations an approach that is based on 
  modeling terms by graphs with structure constraints
  has turned out to be fruitful. 
  In this extended abstract I describe the underlying problems,
  give references, provide examples, indicate the chosen approaches,
  and compare the initial situations as well as the results that have been obtained, and some results that are being developed at present.
\end{abstract}

\hypersetup{pageanchor=true}
\section{Introduction}
  \label{sec:intro}

  For my talk at the 
                     workshop 
                     I prepared two examples from my past and current work 
  that highlight the usefulness and the potential of graph techniques for problems that have
  been approached predominantly as questions about terms:
  increasing sharing in functional programs, 
  and tackling problems about Milner's process semantics for regular expressions. 
  The unifying element of these two illustrations consists in
  modeling terms by term graphs or transition graphs with structure
  constraints (higher-order features or labelings with added conditions),
  and in being able to go back and forth between terms and graphs.
  
  The first illustration, which I only touched on in my talk, concerns the definition, and the efficient implementation
  of maximal sharing for the higher-order terms in the \lambdacalculus\ with letrec.
  For solving this problem,
  Jan Rochel and I developed a representation pipeline from terms via higher-order term graphs
  and first-order term graphs to deterministic finite-state automata.
  
  The setting for the second illustration, on which I focused in my presentation,
  is Milner's process semantics of regular expressions, which yields nondeterministic finite-state
  automata (NFAs) whose equality is studied under bisimilarity. 
  In my current work with Wan Fokkink, I use labelings of process graphs that
  witness direct expressibility by a regular expression via a condition on the graph topology. 
  
  My motivation for explaining these two cases together developed as follows.
  While working on problems concerning the process semantics of regular expressions
  I have repeatedly benefited from the previous work on modeling cyclic \lambdaterms\ by \structureconstrained\ term graphs. 
  It turned out that many concepts and methods that Jan Rochel and I had developed 
  could be adapted in order to define \structureconstrained\ process graphs that directly represent regular expressions
  under the process semantics. 
  It seemed worthwhile to compare the settings and the results
  so that the flow of ideas from one setting to the other, and probably back, might become clearer.
  Perhaps this can be of help in similar situations.
  
  In this extended abstract I explain the setting and the background of the underlying problems,
  provide references, give examples, and informally describe the chosen approaches:
  in Section~\ref{sec:maxsharing}, for the implementation of maximal sharing of functional programs,
  and in Section~\ref{sec:procsem}, for the problems concerning the process semantics of regular expressions.
  In order to highlight differences, and to identify similarities
  that enabled a transferal of ideas between the two illustrations,
  I compare them in Section~\ref{sec:comparison} with respect to 
            the initial situation, the desired concepts, 
  and the defined structure-constrained graphs.


\section{Maximal sharing of functional programs}
  \label{sec:maxsharing}
%
The first example concerns the definition, and the efficient implementation of maximal sharing for functional programs,
and more specifically, for the higher-order terms in the \lambdacalculus\ with \txtletrec\ \cite{grab:roch:2014:ICFP}. 

Graph representations of terms in the \lambdacalculus\ with \txtletrec\
are crucial for the implementation of functional programming languages,
in particular for facilitating the efficient execution of compiled programs 
in sharing-graph form via graph reduction.  
However, these graph representations were never conceived
as term graph representations that keep their intended meaning under bisimilarity. 
In fact they do not behave well under bisimilarity with respect to the unfolding semantics
of terms in the \lambdacalculus\ with \txtletrec. 
In order to study the compactification of functional programs (in their usual language),
Jan Rochel and I therefore looked for 
term graph representations that support compactification under bisimilarity
while preserving the intended meaning, and being easy to compute and to translate back into terms.
Our focus on these desiderata (see also Figure~\ref{fig:motiv:results:lambdaletreccal} later)
led us to structure-constrained term graph representations,
for which we investigated a number of different options \cite{grab:roch:2013:TERMGRAPH}.
We eventually defined classes of `$\lambda$-higher-order-term-graphs' and of `$\lambda$-term-graphs' 
that are closed under functional bisimilarity and have natural correspondences with the terms in the \lambdacalculus\ with \txtletrec\
(see again in~Figure~\ref{fig:motiv:results:lambdaletreccal}).

On this basis Jan Rochel and I 
              developed a `representation pipeline'
from higher-order terms to deterministic finite-state automata (DFAs):
(1)~Terms in the \lambdacalculus\ with \txtletrec\
    can be represented by appropriately defined higher-order term graphs,
    which are first-order term graphs together with higher-order features
    such as a scope function, or an abstraction prefix function, that are defined on the set of vertices (see \cite{grab:roch:2013:TERMGRAPH});
(2)~higher-order term graphs are encoded as first-order term graphs (see also \cite{grab:roch:2013:TERMGRAPH}), and
(3)~first-order term graphs are represented as DFAs (see \cite{grab:roch:2014:ICFP}).
In this way unfolding equivalence on terms is represented by bisimulation equivalence on term graphs (higher-order and first-order),
and ultimately, by language equivalence of DFAs.
In \cite{grab:roch:2014:ICFP} we also define a readback operation from DFAs that arise by the representation pipeline
back to terms in the \lambdacalculus\ with \txtletrec.
This operation makes it possible to go back and forth between terms and representing DFAs: 
it has the property that the representation via (1), (2), and (3) is the inverse of the readback operation.

\begin{figure}[p!]
  \input{figs/pipeline-1.tex}
  \caption{\label{fig:pipeline-1}%
           Stepwise translation of the term 
           $\;\,\labs{\avar}{\labs{f}{\letin{r = \lapp{\lapp{f}{r}}{\avar}}{r}}}\;\,$ in the \lambdacalculus\ with \txtletrec\
           via the construction of its syntax tree, and its modification into a first-order term graph with scope sets
           to obtain a $\lambda$\nb-higher-order term graph in one of two versions: 
           a higher-order term graph
           with scope sets for abstraction nodes, and with an abstraction-prefix function on the set of vertices.}  
\end{figure}
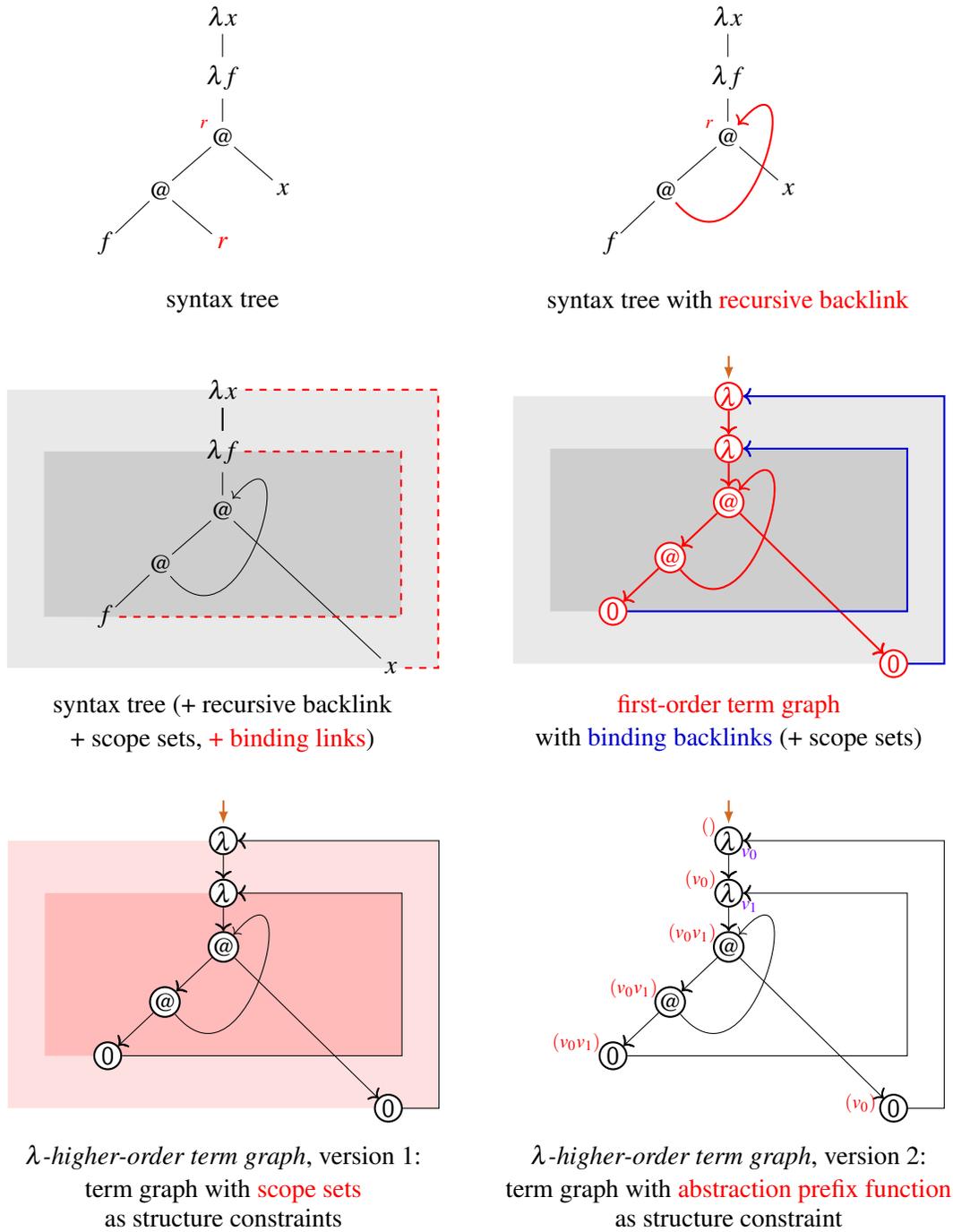

\begin{figure}[p!]
  \input{figs/pipeline-2.tex}
  \vspace*{-2.75ex}
  \caption{\label{fig:pipeline-2}
           Stepwise translation of the term 
           $\;\,\labs{\avar}{\labs{f}{\letin{r = \lapp{\lapp{f}{r}}{\avar}}{r}}}\;\,$ in the \lambdacalculus\ with \txtletrec\
           from the $\lambda$-higher-order term graph obtained in Fig.~\ref{fig:pipeline-1}
           via a $\lambda$\nb-term graph (a first-order term graph in which ends of scopes are encoded by scope vertices)
           and via   an incomplete $\lambda$\nb-DFA
           into 
                a $\lambda$\nb-DFA.
           In the last step a non-accepting state is added to the incomplete $\lambda$\nb-DFA
           to which all missing transitions are directed.%
                }
\end{figure}


Figure~\ref{fig:pipeline-1} and Figure~\ref{fig:pipeline-2} provide an example
for the translation of a term in the \lambdacalculus\ with \txtletrec\
into higher-order and first-order graph representations, and eventually to a finite-state automaton.
Figure~\ref{fig:pipeline-1} covers the part from the syntax tree to $\lambda$\nb-higher-order-term-graphs,
and Figure~\ref{fig:pipeline-2} the remaining part via a $\lambda$\nb-term-graph 
and an `incomplete $\lambda$\nb-DFA' to a `$\lambda$\nb-DFA'.

In Figure~\ref{fig:pipeline-1} we start from the syntax tree of the term, 
model the recursive definition by a recursive backlink,
replace variable names by nameless dummies that have binding backlinks to the corresponding abstraction vertices,
and draw scopes. 
In this way we obtain first-order term graphs with scope sets
that satisfy the conditions for scope sets in the concept of `higher-order term graph' by Blom~\cite{blom:2001}.
We call the specific version obtained here a $\lambda$\nb-higher-order term graph \emph{with scope sets}.
In doing so we distinguish it from a $\lambda$\nb-higher-order term graph \emph{with an abstraction prefix function},
where scopes of abstraction vertices are recorded per vertex $\avert$ via the stack of those abstraction vertices in whose scope $\avert$ resides.
See both versions of $\lambda$\nb-higher-order term graph for the example here at the bottom of Figure~\ref{fig:pipeline-1}.

In Figure~\ref{fig:pipeline-2} we start from the $\lambda$\nb-higher-order-term-graph obtained in Figure~\ref{fig:pipeline-1},
and crucially encode all scope information (recorded by the scope set, or by the abstraction prefix function)
by introducing a scope vertex for the single edge in this example that crosses the boundary of a scope. 
We call the resulting first-order term graph a \emph{$\lambda$\nb-term-graph}.  
By using an intuitive correspondence of term graphs with DFAs, we translate this first-order term graph
further to obtain an incomplete $\lambda$\nb-DFA and eventually a $\lambda$\nb-DFA, both of which represent the 
term $\,\labs{\avar}{\labs{f}{\letin{r = \lapp{\lapp{f}{r}}{\avar}}{r}}}\,$ from which we started.

Via the correspondence statements on which the representation pipeline is based,
unfolding equivalence of terms in the \lambdacalculus\ with \txtletrec\ can be computed in pseudo-quadratic time
$\bigO{n^2 \cdot \fap{\alpha}{n}}$ where $\alpha$ is the inverse Ackermann function (see \cite{grab:roch:2014:ICFP}).  
Again via the correspondences described above, via DFA-minimization, and via the readback
a maximally shared form of higher-order terms can be computed in $\bigO{n^2 \cdot \log n}$ time
(again see \cite{grab:roch:2014:ICFP}).

In order to demonstrate the maximal-sharing method as a manageable optimizing transformation for the compilation of functional programs,
we developed the software tool \cite{roch:grab:2014:maxsharing:tool} that is available on Haskell's Hackage platform. 
Following the definition of maximally shared representations via the representation pipeline in \cite{grab:roch:2014:ICFP}
(see also Rochel's thesis \cite{roch:2016} for more context),
this tool transforms a given functional program in the \lambdacalculus\ with \txtletrec\ 
(the basis of the Core Language of the Glasgow Haskell Compiler) 
into a $\lambda$\nb-term-graph, and then into a $\lambda$\nb-DFA .  
It prints intermediate representations textually, and displays the obtained incomplete $\lambda$\nb-DFA graphically.
The $\lambda$\nb-DFA is then minimized, and 
                        a maximally shared representation of the original program is computed by the readback operation as the result.

Together with Vincent van Oostrom, I have set out to generalize this 
technique of representing higher-order terms as term graphs with added features
that are needed for modeling scopes of binding constructs.
But rather than capturing the constraints on the term graph structure by `ad hoc' features,
we now used `nesting' as the single added structuring concept.
In \cite{grab:oost:2015} we defined, and investigated the behavioral semantics of `nested term graphs' that arise as follows:
by nesting first-order term graphs into the vertices of, initially, a first-order term graph,
and then of nested term graphs that have already been~formed.



\section
        {Process semantics of regular expressions}
        \label{sec:procsem}

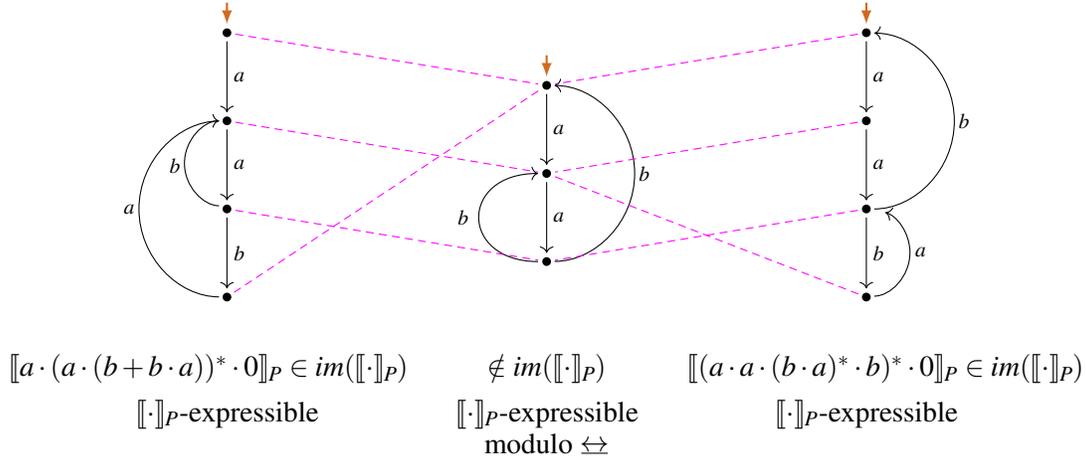
\begin{figure}[p!]
  \vspace*{0.5ex}
  \begin{center}
    \input{figs/expressible.tex}
  \end{center}
  \vspace*{-2.5ex}
  \caption{\label{fig:expressible:expressible-mod-bisim}%
           Process graphs that are expressible by regular expressions via the process semantics~$\protect\procsem{\cdot}$, 
           and expressible modulo bisimilarity~$\protect\sbisim$.
           The graph on the left is the process semantic of 
           \protect\mbox{$a\cdot{(a\cdot(b + b\cdot a))^*} \cdot 0$}, 
           the one on the right of 
           $(a \cdot a \cdot {(b \cdot a)^*} \cdot b)^* \cdot 0$. 
           These graphs are bisimilar, as shown here via bisimulations with
           their bisimulation collapse, a process graph that is not the process semantic of a regular expression.}
\end{figure}

The second illustration concerns the process semantics of regular expressions.
Milner developed a complete axiomatization of bisimulation equivalence for finite process graphs
represented in \muterm\ notation \cite{miln:1984} (1984).
On this basis he turned to descriptions of finite process graphs by regular expressions
with a unary star operation.%
  \footnote{While regular expressions with a binary star operation were introduced by Kleene in \cite{klee:1951} (1951),
            regular expressions with a unary star operation seem to have been first formulated by Copi, Elgot, and Wright \cite{copi:elgot:wrig:1958} (1958).}
Also in \cite{miln:1984} he defined a semantics~$\procsem{\cdot}$ for regular expressions as finite-state processes:
$0$ is interpreted as the deadlock process, $1$ as the immediately terminating process,
letters as actions that lead to termination, and the symbols `$+$', `$\cdot$', and $(\cdot)^*$
as operators that enable choice between processes, sequential composition of processes, and iteration of a process, respectively.
See Figure~\ref{fig:expressible:expressible-mod-bisim}
for two examples of process interpretations of regular expressions via $\procsem{\cdot}$.
Formally, Milner's definition of $\procsem{\cdot}$ yields finite process graphs 
by an inductive definition on the structure of~regular~expressions.

\begin{figure}[tp  ]
\begin{gather*}
 \begin{aligned}
   &
   \AxiomC{\phantom{$\terminates{\stexpone}$}}
   \UnaryInfC{$\terminates{\stexpone}$}
   \DisplayProof
   & \hspace*{-1.5ex} &
   \AxiomC{$ \terminates{\astexpi{1}} $}
   \UnaryInfC{$ \terminates{(\stexpsum{\astexpi{1}}{\astexpi{2}})} $}
   & \hspace*{2ex} &
   \AxiomC{$ \terminates{\astexpi{i}} $}
   \UnaryInfC{$ \terminates{(\stexpsum{\astexpi{1}}{\astexpi{2}})} $}
   \DisplayProof
   & \hspace*{2ex} &
   \AxiomC{$\terminates{\astexpi{1}}$}
   \AxiomC{$\terminates{\astexpi{2}}$}
   \BinaryInfC{$\terminates{(\stexpprod{\astexpi{1}}{\astexpi{2}})}$}
   \DisplayProof
   & \hspace*{2ex} &
   \AxiomC{$\phantom{\terminates{\stexpit{\astexp}}}$}
   \UnaryInfC{$\terminates{(\stexpit{\astexp})}$}
   \DisplayProof
 \end{aligned} 
 \\[1ex]
 \begin{aligned}
   & 
   \AxiomC{$\phantom{a \:\lt{a}\: \stexpone}$}
   \UnaryInfC{$a \:\lt{a}\: \stexpone$}
   \DisplayProof
   & &
   \AxiomC{$ \astexpi{i} \:\lt{a}\: \astexpacci{i} $}
   \UnaryInfC{$ \stexpsum{\astexpi{1}}{\astexpi{2}} \:\lt{a}\: \astexpacci{i} $}
   \DisplayProof 
   & &
   \AxiomC{$ \astexpi{1} \:\lt{a}\: \astexpacci{1} $}
   \UnaryInfC{$ \stexpprod{\astexpi{1}}{\astexpi{2}} \:\lt{a}\: \stexpprod{\astexpacci{1}}{\astexpi{2}} $}
   \DisplayProof
   & &
   \AxiomC{$\terminates{\astexpi{1}}$}
   \AxiomC{$ \astexpi{2} \:\lt{a}\: \astexpacci{2} $}
   \BinaryInfC{$ \stexpprod{\astexpi{1}}{\astexpi{2}} \:\lt{a}\: \astexpacci{2} $}
   \DisplayProof
   & &
   \AxiomC{$\astexp \:\lt{a}\: \astexpacc$}
   \UnaryInfC{$\stexpit{\astexp} \:\lt{a}\: \stexpprod{\astexpacc}{\stexpit{\astexp}}$}
   \DisplayProof
 \end{aligned}
\end{gather*}  
  \vspace*{-2ex}
  \caption{\label{fig:StExpTSS}%
    Transition system specification $\StExpTSS$ of computations enabled by regular expressions.}
\end{figure}

A close variant $\procsemTSS{\cdot}$ of Milner's process semantics~$\procsem{\cdot}$ has later been defined via a transition system specification (TSS):
the TSS~$\StExpTSS$ in Figure~\ref{fig:StExpTSS} explains the operational behavior of 
a regular expression (the option to do a labeled step, or to terminate) inductively for each of the
constants and letters, and for each of the operators.
This TSS is an adaptation for regular expressions with a unary star operation 
of a TSS that was formulated for regular expressions with a binary star operation 
by Bergstra, Bethke, and Ponse \cite{berg:beth:pons:1994} (1994).
By means of the TSS~$\StExpTSS$ the set~$\RegExpsover{\actions}$ of regular expressions over a given set~$\actions$ of action labels 
is endowed with the structure of a labeled transition system (LTS)~$\LTSof{\StExpTSS}$:
there is an $\aact$\nb-transition from $\astexpi{1}$ to $\astexpi{2}$ in $\LTSof{\StExpTSS}$ if and only if $\astexpi{1} \lt{\aact} \astexpi{2}$ is provable in $\StExpTSS$.
Then the variant process interpretation~$\procsemTSS{\astexp}$ of a regular expression~$\astexp$ 
is defined within this encompassing LTS~$\LTSof{\StExpTSS}$ on $\RegExpsover{\actions}$
as the LTS, or process graph, that consists 
                                            of the part of $\LTSof{\StExpTSS}$ that is reachable from $\astexp$. 
This process graph~$\procsemTSS{\astexp}$ can be shown to be finite for every regular expression~$\astexp$.
It 
is closely related,
and in fact always bisimilar to the interpretation $\procsem{\astexp}$ of $\astexp$ according to Milner's process semantics~$\procsem{\cdot}$. 

Every labeled transition system with a finite set of vertices can be construed as a non-deterministic finite-state automaton (NFA).
Therefore the process semantics~$\procsem{\cdot}$ for regular expressions can be viewed as a translation
into 
     NFAs whose equality is studied with respect to bisimilarity, rather than with respect to language equivalence. 
Indeed, Antimirov \cite{anti:1996} (1996) arrived at the same automaton-translation for regular expressions, 
but without process theory and bisimulation equivalence in mind.
He pursued the goal of obtaining for a given regular expression $\astexp$, in a natural way,
an NFA that accepts the language $\fap{L}{\astexp}$ denoted by $\astexp$, and that is smaller than 
NFAs accepting $\fap{L}{\astexp}$ that are obtained by classical algorithms for the translation of regular expressions into NFAs.
For this purpose he introduced, for regular expressions $\astexp\in\RegExpsover{\actions}$
the set of `partial derivatives'~$\partderivs{\aact}{\astexp}$ of $\astexp$ with respect to letters~$\aact\in\actions$,
and a termination predicate $\terms{\astexp}$. More precisely,
he gave definitions by induction on the structure of regular expressions for the functions:
\begin{align*}
  \partderivs{\cdot}{\cdot}
    \funin 
      {\actions} \times \RegExpsover{\actions} & \longrightarrow \powersetof{\RegExpsover{\actions}} 
  &
  \sterms 
    \funin     
      \RegExpsover{\actions} & \longrightarrow \setexp{0,1}\subseteq\nat    
  \\
  \pair{\aact}{\astexp}
    & \longmapsto
       \partderivs{\aact}{\astexp} \punc{,}
  &
  \astexp
    & \longmapsto
        \terms{\astexp} \punc{.}
\end{align*}      
in such a way that the following correspondences hold with respect to the transition system~$\StExpTSS$:
\begin{align*}
  \partderivs{\aact}{\astexp}
    & =
      \descsetexpbig{ \astexpacc\in\RegExpsover{\actions} }
                    { \derivablein{\StExpTSS} \astexp \lt{\aact} \astexpacc } \punc{,}
  &
  \terms{\astexp}
    & =
      \begin{cases}
        \, 1  & \text{ if $\,\derivablein{\StExpTSS} \terminates{\astexp}\;$,}
        \\
        \, 0  & \text{ otherwise$\,$.}
      \end{cases}
\end{align*}
In this way the NFA that is obtained by repeated applications of Antimirov's partial derivatives to a regular expression~$\astexp$
coincides with the NFA that corresponds to the LTS~$\procsemTSS{\astexp}$ as obtained by the TSS~$\StExpTSS$.
That NFA is in turn bisimilar (as a consequence of bisimilarity of the LTSs involved as mentioned above)
to the NFA that corresponds to the interpretation $\procsem{\astexp}$ of $\astexp$ in Milner's process semantics.

Unlike for the standard language semantics~$\langsem{\cdot}\,$, 
not every NFA can be expressed by a regular expression under the process interpretation~$\procsem{\cdot}\,$.
That is, not every NFA is 
         bisimilar to the process translation NFA of some regular expression.
This is witnessed by the two examples in Figure~\ref{fig:not-expressible-mod-bisim}, 
both of which were suggested already by Milner.
He showed, in \cite{miln:1984}, that the three-vertex example without termination in Figure~\ref{fig:not-expressible-mod-bisim} is not \procsemexpressible.
That the second example with two termination-permitting vertices in Figure~\ref{fig:not-expressible-mod-bisim} is not \procsemexpressible\ 
was proved by Bosscher~\cite{boss:1997}. 

\begin{figure}[p!]
  \vspace*{0.5ex}
  \begin{center}
    \input{figs/not-expressible-mod-bisim.tex}
  \end{center}
  \vspace*{-2.5ex}
  \caption{\label{fig:not-expressible-mod-bisim}%
           Process graphs that are neither \procsemexpressible\ nor \procsemexpressible\ modulo bisimilarity~$\sbisim$.
           In the process graph on the left, both vertices permit immediate termination (indicated by the outer circles).%
           }  
\end{figure}

%
\begin{figure}[p!]
  %
  %
  \vspace*{-1.25ex}
  \begin{align*}
    {({B}_1)}   & & \hspace*{-4ex}
               \stexpsum{\astexp}{(\stexpsum{\bstexp}{\cstexp})} & = \stexpsum{(\stexpsum{\astexp}{\bstexp})}{\cstexp}  & \hspace*{4ex}
       {({B}_7)}  & &   \hspace*{-4ex}
                   \stexpprod{\astexp}{\stexpit{\stexpzero}} & = \astexp \hspace*{-4.5ex}
                                                             \\  
    {({B}_2)}   & & \hspace*{-4ex}
               \stexpprod{(\stexpprod{\astexp}{\bstexp})}{\cstexp}
                 & = \stexpprod{\astexp}{(\stexpprod{\bstexp}{\cstexp})} & 
       \textcolor{red}{({B}_8)}  & &    \hspace*{-4ex}
                    \textcolor{red}{\stexpprod{\astexp}{\stexpzero}} & \mathrel{\textcolor{red}{=}} \textcolor{red}{\stexpzero}
                                                             \\
    {({B}_3)}   & & \hspace*{-4ex}
               \stexpsum{\astexp}{\bstexp} & = \stexpsum{\bstexp}{\astexp}   &
       {({B}_9)}  & &     \hspace*{-4ex}
                      \stexpsum{\astexp}{\stexpzero} & = \astexp
                                                             \\   
    {({B}_4)}   & &  \hspace*{-4ex}
                \stexpprod{(\stexpsum{\astexp}{\bstexp})}{\cstexp}
                  & = \stexpsum{\stexpprod{\astexp}{\cstexp}}{\stexpprod{\bstexp}{\cstexp}} & 
       {({B}_{10})}  & &  \hspace*{-4ex}
                   \stexpit{\astexp} & = \stexpsum{\stexpit{\stexpzero}}{\stexpprod{\astexp}{\stexpit{\astexp}}}
                                                             \\ 
    \textcolor{red}{({B}_5)}   & &  \hspace*{-4ex}    
                \textcolor{red}{\stexpprod{\astexp}{(\stexpsum{\bstexp}{\cstexp})}} 
                  & \mathrel{\textcolor{red}{=}} \textcolor{red}{\stexpsum{\stexpprod{\astexp}{\bstexp}}{\stexpprod{\astexp}{\cstexp}}}   &
       {({B}_{11})}  & &   \hspace*{-4ex}
                    \stexpit{\astexp} & = \stexpit{(\stexpsum{\stexpit{\stexpzero}}{\astexp})}
                                                             \\ 
    {({B}_6)}  & &  \hspace*{-4ex}
               \stexpsum{\astexp}{\astexp} & = \astexp  &
       \mediumblue{({A}_8)} & &   \hspace*{-4ex}
               \mediumblue{\stexpprod{\stexpzero}{\astexp}} & \mathrel{\mediumblue{=}} \mediumblue{\stexpzero}  
                                                             \\ 
       {(\REFL)} & &   \hspace*{-4ex}
               \astexp & = \astexp
  \end{align*}
  %
  \vspace*{-4ex}
\begin{alignat*}{4}
  \begin{aligned}[c]
    \AxiomC{$\astexp = \bstexp$}
    \RightLabel{\SYMM}
    \UnaryInfC{$\bstexp = \astexp$}
    \DisplayProof
  \end{aligned}  
    & \quad\;\, & 
  \begin{aligned}[c]
    \AxiomC{$\astexp = \bstexp$}
    \AxiomC{$\bstexp = \cstexp$}
    \RightLabel{\TRANS}
    \BinaryInfC{$\astexp = \cstexp$}
    \DisplayProof
  \end{aligned} 
    & \quad\;\, &  
  \begin{aligned}
    \AxiomC{$\astexp = \bstexp$}
    \RightLabel{\CXT}
    \UnaryInfC{$\acxtap{\astexp} = \acxtap{\bstexp}$}
    \DisplayProof
  \end{aligned}
    & \quad\;\, & 
  \begin{aligned}[c]  
    \AxiomC{$ \astexp  =  \stexpsum{\stexpprod{\bstexp}{\astexp}}{\cstexp} $}
    \RightLabel{\parbox{\widthof{$\FIX$ {\scriptsize (if $\bstexp$ does not}}}
                       {\rule{0pt}{3.35ex}%
                        $\FIX$
                          {\scriptsize (if $\bstexp$ does not}%
                              \\[-1ex]
                              \phantom{$\FIX$ }
                          {\scriptsize have e.w.p.)}}}
    \UnaryInfC{$ \astexp = \stexpprod{\stexpit{\bstexp}}{\cstexp} $}
    \DisplayProof
  \end{aligned}
\end{alignat*} 
  \vspace*{-3.25ex}
  \caption{\label{fig:aanderaa:2:milner}%
           Complete axiomatization of equality of regular expressions under the \protect\underline{\protect\smash{language semantics}} $\langsem{\cdot}\,$.
           The system is due to Aanderaa, and corresponds to Salomaa's system by commuting product expressions. 
           Axiom $\mediumblue{({A}_8)}$ from Salomaa's system is derivable, and not part of Aanderaa's system.   
           Axioms that are not sound under the process semantics $\procsem{\cdot}$ are colored~in~red.
           Milner's axiomatization \BPAstarzeroone\ of bisimilarity of regular expressions under the \protect\underline{\protect\smash{process semantics}} $\procsem{\cdot}$
           arises by dropping the unsound axioms (in red).}
\end{figure}
%

Still in \cite{miln:1984}, 
Milner adapted the complete axiomatization by Salomaa \cite{salo:1966} for language equivalence of regular expressions.
He started from a version of Salomaa's system in which
all product expressions in the axioms and rules are commuted, see Figure~\ref{fig:aanderaa:2:milner}.
The rule $\FIX$ is subject to the `non-algebraic' side-condition that the regular expression $\astexp$ does not have the 
`empty word property', that is, the language interpretation $\langsem{\astexp}$ of $\astexp$
does not contain the empty word.
This system is close to the complete axiomatization for language equivalence
that was presented by Aanderaa \cite{aand:1965} independently from Salomaa's work (Aanderaa's system was probably not directly known to Milner).
Milner dropped the two rules from the system that are unsound under the process semantics
(left-distributivity ${B}_5$, and the axiom ${B}_8$),
but additionally took up the axiom~$(A_{8})$ from Salomaa's original system, 
which describes a correct interaction property of $\szero$ as deadlock with process concatenation. 
The resulting system is sound for the process semantics~$\procsem{\cdot}\,$.
It has later been called \BPAstarzeroone\ 
as an adaptation of Basic Process Algebra \BPA\ to regular expressions as terms that describe process behavior with respect to $\procsem{\cdot}$.

Milner noticed that completeness for \BPAstarzeroone\ cannot be settled directly by Salomaa's arguments. 
This is due to the incompleteness modulo bisimilarity~$\sbisim$ of the image of the process semantics~$\procsem{\cdot}\,$.
That namely implies that not every finite regular system of equations is solvable by a regular expression
(for example, specifications that correspond to the process graphs in Figure~\ref{fig:not-expressible-mod-bisim} are not solvable).
However, being able to solve arbitrary finite regular systems of equations by regular expressions 
is a crucial lemma in  Salomaa's and Aanderaa's completeness proofs.
Recognizing this difficulty, Milner formulated the question as to whether \BPAstarzeroone\ is indeed a complete axiomatization
for bisimilarity of interpretations of regular expressions in the process semantics~$\procsem{\cdot}$. 
In addition, he also formulated the problem of characterizing those process graphs that are bisimilar to process interpretations of regular expressions,
and a star-height~problem for regular expressions over a single-letter alphabet.
  
The known approaches to these questions by Milner fall, broadly speaking, into two groups
that are distinguished by how they model processes that are represented by regular expressions:
either by working with process terms whose operational semantics is governed by structural operational semantics (SOS) rules,
such as TSSs,
or by reasoning about regular recursive process specifications of a certain structure. 
Taking a new approach, I have set out to use structure-constrained process graphs, see below.     

Building on work from the process term tradition,  
Fokkink (1996-97) showed that the restriction of Milner's system to exit-less iteration,
which he called `\perpetualloop' and `terminal cycle', 
is complete for the general case with `empty' 1-steps \cite{fokk:1996:terminal:cycle:LGPS},
and for the easier case without \cite{fokk:1997:pl:ICALP}. 
To achieve this result
he completely overturned Salomaa's and Aanderaa's proof technique 
of extension of terms (obtaining a common extension for semantically equal terms) 
into its contrary, a strategy of \mbox{term minimization}.   

Also working with term calculi for process terms, 
Corradini, De~Nicola, and Labella \cite{corr:nico:labe:2002}
define a subclass of regular expressions, those without occurrences of $\szero$ that satisfy the `hereditary non-empty word property (hnewp)',
and give a `(purely) equational' axiomatization for $\procsem{\cdot}$ on regular expressions with these restrictions. 
Indeed their result shows that Milner's axiomatization without the axioms involving $\szero$
is complete for regular expressions from that class.
This is because for regular expressions with hnewp the \nonequational\ side-condition on the fixed-point rule $\FIX$
is irrelevant, and therefore can be dropped, which turns the axiomatization into a purely equational one.

Regular expressions that may contain $\szero$, but satisfy the property hnwep of Corradini, De~Nicola, and Labella can be characterized as follows:
for {\underline{\smash{no}}} iteration subexpression $\stexpit{\bstexp}$ of $\astexp$ 
does $\procsem{\bstexp}$ proceed to a process $p$ such that:
  $p$ has the option to immediately terminate, \underline{\smash{and}}
  $p$ has the option to do a proper step, and terminate later.
Motivated by this, I call these expressions `$\stexpone$-return-less(-under-$*$)'.
They turned out to be relevant in my current work on structure-constrained process graphs, see below.        

Using recursive specifications to formalize processes that are induced by regular expressions,
Baeten and Corradini (2005) introduced `well-behaved specifications' \cite{baet:corr:2005}.
These systems of equations are arranged according to trees with back-bindings (`palm trees') 
with a `loop--exit' structure requirement.
This concept enabled Baeten, Corradini, and myself to show that expressibility modulo bisimilarity of a finite process graph by a regular expression 
is decidable \cite{baet:corr:grab:2007}, although via a super-exponential procedure.

My current approach to the axiomatization problem 
(in work with Wan Fokkink)
takes the conscious step to reasoning about process graphs for which
the palm-tree form is relaxed significantly as constraint. 
A crucial step is the formulation of a concept of transition graph labeling
that is inspired by Milner's notion of `loop'.
Transitions (action-labeled edges) 
are decorated by additional marker labels
that witness that the syntax tree of a regular expression can be inscribed on to a (typically cyclic) process graph.
In this way a labeling witnesses that the process graph can be expressed \emph{directly} by a regular expression.  
This opens the way to develop bisimilarity-preserving transformations
of directly expressible process~graphs, in order to constructively connect any two given directly expressible process graphs
that are bisimilar. 

Figure~\ref{fig:motiv:results:procsem} in Section~\ref{sec:comparison} gathers the initial motivation for defining structure-constrained process graphs,
and puts the desiderata here in the context of the properties of Milner's process semantics~$\procsem{\cdot}$. 
It also gives a preliminary overview on results that are being developed at the moment.

%

\begin{figure}[p!]
  \begin{center}
    \input{figs/LEE-extraction-2-LEE.tex}
  \end{center}
  \vspace*{-0.75ex}
  \caption{\label{fig:loop:elimination:synthesis}%
           Loop elimination for the left process graph in the upper row by repeatedly identifying a \loopentry\  transition,
           then removing it, and performing garbage collection. 
           Since a process graph without infinite behavior is reached,
           the original process graph has the property \LEE.
           In the second row a structured version of the original graph is reassembled in converse direction 
           by using the eliminated~loops.
           }
\end{figure}
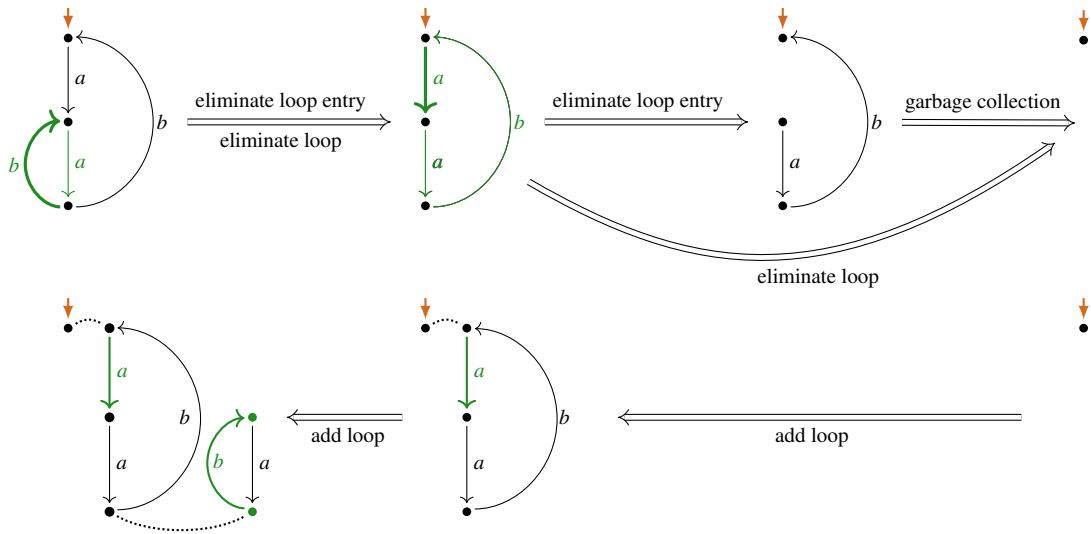  

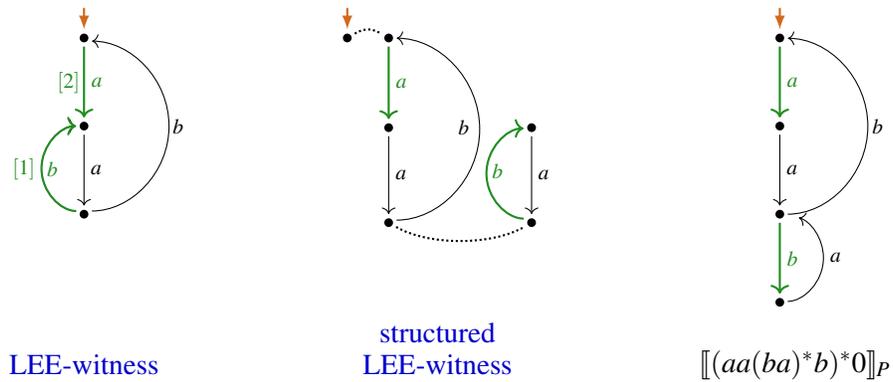
\begin{figure}[p!]
  \begin{center}
    \input{figs/LEE-extraction-2-result.tex}
  \end{center}
  \vspace*{-1.5ex}
  \caption{\label{fig:LEE-witness}%
           A \LEEwitness\ for the original process graph in Fig.~\ref{fig:loop:elimination:synthesis}
           is obtained by overlaying the loops from the structured version that has been obtained
           by loop-addition synthesis in Fig.~\ref{fig:loop:elimination:synthesis},
           and by number-labels that record the order of loop removal.
           The structured form of the \LEEwitness\ indicates a correspondence
           with the process semantics of one of the regular expressions considered in Fig.~\ref{fig:expressible:expressible-mod-bisim}.%
           }
\end{figure}

By modifying a concept introduced by Milner in \cite{miln:1984},
we call a process graph a `loop' if all paths from the start vertex return to it, and termination is only permitted at the start vertex.
A `loop subgraph' in a process graph $G$ is a loop that is generated from a vertex $\avert$ of $G$ 
by a set $T$ of `\loopentry\ transitions' from $\avert$ as follows:
the subgraph of $G$ that consists of all vertices and edges that are reachable on paths departing from $\avert$
via an edge in $T$ until $\avert$ is reached again. 
Furthermore we call `loop elimination' a procedure that, starting from a given process graph
repeatedly identifies a loop subgraph, drops its \loopentry\ transitions, and performs garbage collection
(removing vertices and edges that have become unreachable from the start vertex). 
We say that a process graph $G$ satisfies the \emph{loop existence and elimination condition~(LEE)}
if by loop elimination from $G$ a process graph without an infinite behavior (that is, without an infinite trace) can be reached.  

Figure~\ref{fig:loop:elimination:synthesis} in its upper row shows two loop elimination steps
that are performed starting from the process graph in the middle of Figure~\ref{fig:expressible:expressible-mod-bisim}.
These steps lead to a process graph without any transitions, and hence without an infinite trace.
Thus they witness that the 
                           original process graph has the property \LEE. 
By contrast, none of the two process graphs in Figure~\ref{fig:not-expressible-mod-bisim} contains a loop subgraph:
the two-vertex graph does not because the termination condition of a loop would be violated;
and the three-vertex graph does not because no transition from a vertex $\avert$ generates a subchart in which all infinite paths return to $\avert$.
Hence these process graphs, which are not \procsemexpressible\ modulo~$\sbisim$, 
do not satisfy the property~LEE.

In its lower row, Figure~\ref{fig:loop:elimination:synthesis} records a procedure of reassembly 
of the process graph in the upper left corner from the results that have been obtained during loop elimination.
Thereby an approximation of the original process graph is assembled that is structured by \onetransitions.
We call it a \emph{structured \LEEwitness}. 
Figure~\ref{fig:LEE-witness} indicates that a \emph{\LEEwitness} is obtained from the structured version 
by overlaying the separately recorded loop subgraphs on to the original process graph,
and by labeling the identified \loopentry\  transitions according to the order in which they have been removed
during loop elimination. 

A \LEEwitness\ records the loop elimination procedure in a process graph by
marking transitions that have been recognized as \loopentry\  transitions 
with a label that indicates its number (or nesting depth) in the procedure.
It is subject to conditions that follow from this intuition, and the requirement
that loop elimination leads to a process graph without an infinite trace.
Thus a \LEEwitness\ is a labeling of a process graph that is subject to appropriate conditions
that witnesses that the graph satisfies \LEE.
In this way we obtain a class of \structureconstrained\ process graphs
that consists of all graphs that have a \LEEwitness, and hence satisfy \LEE. 
The arising class properly extends the class of process graphs that are the process semantics of some regular expression:
the process graph in the middle of Figure~\ref{fig:expressible:expressible-mod-bisim} 
has a \LEEwitness, and satisfies \LEE\ (see Figure~\ref{fig:loop:elimination:synthesis} and Figure~\ref{fig:LEE-witness}),
but it is not \procsemexpressible.

The concept of \LEEwitness\ is an important technical tool 
for investigating transformations between process graphs that satisfy the graph-topological property~\LEE,
and for extracting regular expressions from such process graphs. 
It facilitates a number of results such as the following:
(1)~\LEE\ is preserved under functional bisimilarity~$\sfunbisim$ for process graphs without empty steps.
    The proof of this statement relies on the fact that \LEEwitnesses\ can be transferred along functional bisimulations. 
(2)~From every process graph~$G$ without \onetransitions\ that satisfies \LEE\ 
    a \txtonereturnless\ regular expression $\astexp$ can be extracted 
    for which $\procsem{\astexp} \bisim G$ holds,
    that is, such that $\astexp$ expresses $G$ under $\procsem{\cdot}$ modulo bisimilarity.
    This statement can be proved by using the number labels of the \loopentry\ transitions in a \LEEwitness\
    to define a bottom-up extraction procedure of a regular expression.
    
These statements lead to 
a new partial answer to Milner's question about how \procsemexpressibility\ of finite process graphs can be characterized:
A finite process graph $G$ is \procsemexpressible\ by a {\txtonereturnless} regular expression
if and only if the bisimulation collapse of $G$ satisfies the property \LEE.

\section{Comparison desiderata and results}
  \label{sec:comparison}

\begin{figure}[bp!]
  %
\begin{description}
  \item[\framebox{{\it \lambdacalculus\ with \txtletrec} \text{\rm with respect to the} {\it unfolding semantics}}] 
    \mbox{}
    \begin{description}
      \item[{\it Known:}]
        graph representations of terms in the \lambdacalculus\ with \txtletrec\ are used in compilers of functional languages. 
        However:
        \begin{itemize}
          \item
            these graph representations were not intended for use under transformations that involve bisimilarity~$\sbisim$,
            and do not behave well under such transformations.
        \end{itemize}
        
      
      \item[{\it Aim:}]
        a term graph semantics that:
        \begin{itemize}[label=$\triangleright$]
         \item
           has a natural correspondence with terms in \lambdacalculus\ with \txtletrec, 
         \item
           supports compactification under bisimilarity $\sbisim$,
         \item  
           permits efficient operations to translate between terms to graphs.
        \end{itemize}
        
      \item[{\it Defined:}] \emph{Structure-constrained term graphs} as a semantics for terms in the \lambdacalculus\ with \txtletrec:
        \begin{itemize}[label=$\myblacktriangleright$,itemsep=0.25ex]\vspace*{0.25ex}
          \item
            the class $\classlhotgs$ of higher-order $\lambda$\nb-term graphs, with interpretation function $\graphsemC{\classlhotgs}{\cdot}$,
          \item  
            the class $\classltgs$ of first-order $\lambda$\nb-term graphs, with interpretation function $\graphsemC{\classltgs}{\cdot}$.
        \end{itemize}
        They have the following properties:
        \begin{enumerate}[label={(\roman*)},leftmargin=*,align=right,labelsep=0.9ex,itemsep=0.25ex]
          \item
            $\lambda$\nb-term graphs are first-order term graph encodings of $\lambda$\nb-higher-order term graphs,
          \item[{\bf (ii)}]
            $\classlhotgs$ and $\classltgs$ are closed under functional bisimilarity $\sfunbisim\,$ (and hence under collapse),
          \item[{\bf (iii)}]
            there is a back-/forth correspondence with terms in the {\lambdacalculus\ with \txtletrec} such that:
            \begin{itemize}
              \item
                there are efficient translation and readback operations (computable in $O(n^2 \log n)$ and $O(n\log n)$ time),
              \item
                the translation is the inverse of the readback.   
            \end{itemize} 
        \end{enumerate}
    \end{description}
\end{description}
  \vspace*{-1ex}
  \caption{\label{fig:motiv:results:lambdaletreccal}%
           Motivation for developing \protect\structureconstrained\ term graph representations
           for the first illustration, the \protect\lambdacalculus\ with \protect\txtletrec$\,$;
           and an overview of the obtained concepts and results.
           The key results (ii) and (iii) are highlighted
           as they correspond to analogous results for the second illustration,
           see Fig.~\ref{fig:motiv:results:procsem}.}
\end{figure}

Apart from demonstrating the usefulness of working with \structureconstrained\ graphs,
another motivating aim for my talk was 
to obtain a clearer view of the similarity and the difference of the two situations. 
In particular I wanted to understand 
why I was able to benefit from a flow of ideas from the first to the second illustration.
As a first step towards a better understanding I assembled, for each of the two settings, 
a list of the motivations and desiderata for graph representations arising from the initial problems,
and of the results that have been obtained, or that are being developed. 
These overviews are gathered
in Figure~\ref{fig:motiv:results:lambdaletreccal} and in Figure~\ref{fig:motiv:results:procsem}. 

The initial situations are markedly different:
a graph semantics that is studied under bisimilarity is provided by Milner's process semantics of regular expressions,
whereas graph representations for cyclic \lambdaterms\ that are used in compilers do not behave well under bisimilarity.
For representing cyclic \lambdaterms\ an appropriate class of term graph representations needed to be defined,
for example one based on Blom's higher-order term graphs~\cite{blom:2001}.
Yet also the incompleteness under functional bisimilarity of the image of the process semantics
stimulated extending this class of graphs to one with more satisfying properties. 
%

The joining element of the results obtained in the two settings
consists in the definition of classes of \structureconstrained\ graphs
that, on the one hand, are closed under functional bisimilarity (and hence are closed under the operation of taking the bisimulation collapse),
and that, on the other hand, enable a natural, and efficiently computable correspondence with the class of terms that is relevant for the setting.
This observation is highlighted in Figure~\ref{fig:motiv:results:lambdaletreccal} and Figure~\ref{fig:motiv:results:procsem} 
by the items with boldface numbers:
(ii)~for closedness under functional bisimilarity $\sfunbisim$,
and (iii)~for the natural correspondence with terms.

In conclusion I want to repeat a request that I have put to the participants of the workshop:
I am interested in, and would like to hear about, other situations and settings
in which \structureconstrained\ graph representations might be useful,
or have already been developed and used successfully.


\begin{figure}[tp!]
  %
\begin{description}
  \item[\framebox{{\it Regular expressions} \text{\rm with respect to the} {\it process semantics}}] \mbox{}
    \begin{description}
      \item[{\it Given:}]
        Milner's process graph semantics~$\procsem{\cdot}$ was designed for study under bisimilarity $\sbisim$.
        \\
        However, the semantics~$\procsem{\cdot}$ has some peculiar properties:
        \begin{itemize}
          \item
            the image of $\procsem{\cdot}$ is not closed under functional bisimilarity~$\sfunbisim$
          \item
            the image of $\procsem{\cdot}$ is incomplete modulo bisimilarity~$\sbisim$
        \end{itemize}
      \item[{\it Aim:}] 
        in order to tackle completeness of Milner's axiomatization, and the recognizability of \procsemexpressibility\ modulo $\sbisim$,
        it is desirable to:
        \begin{itemize}[label=$\triangleright$]
          \item
            reason with (`sufficiently many') graphs that are \procsemexpressible\ modulo $\sbisim$;
          \item   
            understand incompleteness modulo $\sbisim$ by a structural graph property.
        \end{itemize} 
        
      \item[{\it Defined / under construction / current aim:}]
        \emph{Structure-constrained process graphs}, in particular:
        \begin{itemize}[label=$\myblacktriangleright$,itemsep=0.25ex]\vspace*{0.25ex}
          \item
            the class of finite process graphs with the property \LEE\
            which consists of all those process graphs that have a (layered) \LEEwitness\ labeling.
        \end{itemize}
        It has the following properties:
        \begin{enumerate}[label={(\roman*)},leftmargin=*,align=right,labelsep=0.9ex,itemsep=0.25ex]
          \item[(i)]
            it extends the image of the process semantics $\procsem{\cdot}\,$;
          \item[{\bf (ii)}]
           it is closed under functional bisimilarity~$\sfunbisim\,$ (and hence under bisimulation collapse)
            in the special case of the absence of \onetransitions\ (empty-step transitions);
          \item[{\bf (iii)}]
            it permits efficient back and forth translations to and from \txtonereturnless\ expressions;
          \item[(iv)]
            it characterizes \procsemexpressibility\ modulo~$\sbisim$ by a \mbox{\txtonereturnless} regular expression of a graph's collapse: 
            a finite process graph~$G$ is \procsemexpressible\ modulo~${\sbisim}$ by a \txtonereturnless\ regular expression
              {if and only if}
            the bisimulation collapse of $\aprocgraph$ satisfies~\LEE.
        \end{enumerate}
    \end{description}
\end{description}
  \vspace*{-1ex}
  \caption{\label{fig:motiv:results:procsem}%
           Motivation for developing structure-constrained process graphs
           for the second illustration, the process semantics for regular expressions;
           and an overview of the results that we are currently working out.
           The key results~(ii) and (iii) are emphasized with their labels in boldface
           in order to highlight their correspondence with the analogous results (ii) and (iii) for the first illustration
           in Fig.~\ref{fig:motiv:results:lambdaletreccal}.}
\end{figure}

\enlargethispage{2.5ex}
\paragraph{Acknowledgment.}
  I want to thank Luca Aceto for his detailed comments and for valuable hints at substantial issues,
  Ruben Becker for spotting several errors and inconsistencies,
  Omar Inverso for a good number of concise, acute, and helpful suggestions,
  and Maribel Fernandez for pointing me to some structural improvements.

\bibliographystyle{eptcs}
\bibliography{ext-abstract-pp.bib}

\end{document}

%% file: packages.tex
\usepackage{amsfonts,amsmath,amsthm,amscd,amssymb,float}
\usepackage{afterpage}
\usepackage{breakurl}
\usepackage{bussproofs}
\usepackage{calc}
\usepackage{centernot}
\usepackage{color,graphicx}
\usepackage{enumitem}
\usepackage{hyperref}
\usepackage[section]{placeins}
\usepackage{soul}
\usepackage{stmaryrd}
\usepackage{underscore}           
\usepackage{xspace}
\usepackage{placeins}
\usepackage{tikz-cd}

\usepackage{tikz}
\usetikzlibrary{arrows,arrows.meta,calc,automata,angles,quotes,
                positioning,shapes.geometric,shapes.symbols,shapes.multipart,through,trees,
                decorations.markings,decorations.text
                }

\pgfdeclaredecoration{funbisim}{final}{
  \state{final}[width=\pgfdecoratedpathlength]{
    \draw[->]
      (0,\pgfdecorationsegmentamplitude+0.1mm) -- +(\pgfdecoratedpathlength,0);
    \draw[-]
      (0,-\pgfdecorationsegmentamplitude-0.1mm) -- +(\pgfdecoratedpathlength,0);}}
\tikzset{
  funbisim/.style={
    decoration={funbisim, amplitude=0.25ex},
    decorate,
    funbisim options/.style={#1}    
  }}

\pgfdeclaredecoration{bisim}{final}{
  \state{final}[width=\pgfdecoratedpathlength]{
    \draw[<->]
      (0,\pgfdecorationsegmentamplitude+0.1mm) -- +(\pgfdecoratedpathlength,0);
    \draw[-]
      (0,-\pgfdecorationsegmentamplitude-0.1mm) -- +(\pgfdecoratedpathlength,0);}}
\tikzset{
  bisim/.style={
    decoration={bisim, amplitude=0.25ex},
    decorate,
    bisim options/.style={#1}    
  }}


%% file: macros.tex
\newcommand{\nf}{\normalfont}

\newcommand{\myblacktriangleright}{\raisebox{0.2ex}{\scalebox{0.75}{$\blacktriangleright$}}}

\newcommand{\funin}{\mathrel{:}}
\newcommand{\fap}[2]{{#1}(\hspace*{-0.5pt}{#2}\hspace*{-0.5pt})}

\newcommand{\iap}[2]{#1_{#2}}

\newcommand{\pap}[2]{#1^{#2}}
\newcommand{\bpap}[3]{#1_{#2}^{#3}}
\newcommand{\pbap}[3]{#1_{#3}^{#2}}

\newcommand{\nb}{\nobreakdash}
\newcommand{\punc}[1]{\ensuremath{\hspace*{1.5pt}{#1}}}

\newcommand\tuple[1]{\langle #1 \rangle}

\newcommand\tuplespace{\hspace*{0.5pt}}
\newcommand\pair[2]{\tuple{#1, \tuplespace #2}}

\newcommand{\setexp}[1]{\left\{{#1}\right\}}

\newcommand{\descsetexpbigmid}{\mathrel{\big\vert}}

\newcommand{\descsetexpbig}[2]{\bigl\{{#1}\descsetexpbigmid{#2}\bigr\}}

\newcommand{\spowersetof}{{\cal P}}
\newcommand{\powersetof}{\fap{\spowersetof}}

\newcommand{\nat}{\mathbb{N}}

\newcommand{\acxt}{C}

\newcommand{\cxtap}[2]{{#1}[{#2}]}
\newcommand{\acxtap}{\cxtap{\acxt}}

\newcommand{\sbigO}{O}
\newcommand{\bigO}{\fap{\sbigO}}

\newcommand{\muterm}{{$\mu$}\nb-term}

\newcommand{\perpetual}{per\-pet\-u\-al}
\newcommand{\perpetualloop}{\perpetual-loop}

\newcommand{\slangsem}{\boldsymbol{L}}
\newcommand{\langsem}[1]{\llbracket{#1}\rrbracket_{\hspace*{-0.5pt}\slangsem}}

\newcommand{\sprocsem}{\boldsymbol{P}}
\newcommand{\procsem}[1]{\llbracket{#1}\rrbracket_{\hspace*{-0.5pt}\sprocsem}}  
\newcommand{\procsemTSS}[1]{\llbracket{#1}\rrbracket^{\prime}_{\hspace*{-0.5pt}\sprocsem}}  
 
\newcommand{\procsemexpressible}{$\procsem{\cdot}$\nb-ex\-press\-ible}
\newcommand{\procsemexpressibility}{$\procsem{\cdot}$\nb-ex\-press\-ibi\-li\-ty}

    \newcommand{\sonereturnlessunderstar}
                                         {\text{\bf\textcolor{red}{\st{$\boldsymbol{\stexpone}$r}$\hspace*{-0.5pt}\backslash\hspace*{-1pt}\star$}}}

\newcommand{\aprocgraph}{{G}}

\newcommand{\graphsemC}[2]{\llbracket{#2}\rrbracket_{#1}}
\newcommand{\classlhotgs}{{\cal H}}

\newcommand{\classltgs}{{\cal F}} 

\newcommand{\sLEE}{\text{\nf{LEE}}}
\newcommand{\LEE}{\sLEE}

\newcommand{\LEEwitness}{$\LEE$\nb-wit\-ness}

\newcommand{\LEEwitnesses}{$\LEE$\hspace*{1.25pt}\nb-wit\-nes\-ses}

\newcommand{\txtonereturnless}{1-return-less}

\newcommand{\sterminates}{\downarrow}
\newcommand{\terminates}[1]{{#1}{\sterminates}}

\newcommand{\slt}[1]{{\xrightarrow{#1}}}
\newcommand{\lt}[1]{\mathrel{\slt{#1}}}

\newcommand{\RegExps}{\mathit{RegExp}}
\newcommand{\RegExpsover}{\fap{\RegExps}}

\newcommand{\actions}{A}

\newcommand{\astexp}{e}
\newcommand{\bstexp}{f}
\newcommand{\cstexp}{g}

\newcommand{\astexpi}{\iap{\astexp}}

\newcommand{\astexpacc}{\astexp'}

\newcommand{\astexpacci}{\iap{\astexpacc}}

\newcommand{\szero}{0}
\newcommand{\sone}{1}
\newcommand{\sstar}{*}

\newcommand{\stexpzero}{0}
\newcommand{\stexpone}{1}

\newcommand{\stexpit}[1]{{#1^{\sstar}}}

\newcommand{\sstexpprod}{{\cdot}}
\newcommand{\stexpprod}[2]{{#1}\mathrel{\sstexpprod}{#2}}
\newcommand{\sstexpsum}{+}
\newcommand{\stexpsum}[2]{{#1}\sstexpsum{#2}}

\newcommand{\aact}{a}

\newcommand{\spartderivs}{\partial}
\newcommand{\partderivs}[1]{\fap{\spartderivs_{#1}}}
\newcommand{\sterms}{\textit{tm}}
\newcommand{\terms}{\fap{\sterms}}

\newcommand{\REFL}{\textrm{\nf Ref\/l}}
\newcommand{\SYMM}{\textrm{\nf Symm}}
\newcommand{\TRANS}{\textrm{\nf Trans}}
\newcommand{\CXT}{\textrm{\nf Cxt}}

\newcommand{\FIX}{\textrm{Fix}}

\newcommand{\BPA}{\text{\sf BPA}}
\newcommand{\BPAstarzeroone}{\text{$\pbap{\BPA}{\sstar}{\text{\nf\sf 0},\text{\nf\sf 1}}$}}

\newcommand{\saTSS}{{\cal{T}}}
\newcommand{\StExpTSS}{\text{$\saTSS$}}

\newcommand{\aLTS}{\mathcal{L}}
\newcommand{\LTSof}{\fap{\aLTS}}

\newcommand{\sderivablein}[1]{\vdash_{#1}}
\newcommand{\derivablein}[1]{\sderivablein{#1}}

\newcommand{\letin}[2]{\txtlet\;{#1}\;\txtin\;{#2}}

\newcommand{\txtin}{{\text{\sf in}}}
\newcommand{\txtlet}{{\text{\sf let}}}
\newcommand{\txtletrec}{\text{\normalfont\sf letrec}}

\newcommand{\sfunbisim}{%
    \setbox0=\hbox{\kern-.1ex{$\rightarrow$}\kern-.1ex}
    \setbox1=\vbox{\hbox{\raise .1ex \box0}\hrule}%
    {\hbox{\kern.05ex\box1\kern.1ex}}
  }

\newcommand{\sconvfunbisim}{%
    \setbox0=\hbox{\kern-.1ex{$\leftarrow$}\kern-.1ex}
    \setbox1=\vbox{\hbox{\raise .1ex \box0}\hrule}%
    {\hbox{\kern.05ex\box1\kern.1ex}}
  }

\newcommand{\sbisim}{%
    \setbox0=\hbox{\kern-.1ex{$\leftrightarrow$}\kern-.1ex}
    \setbox1=\vbox{\hbox{\raise .1ex \box0}\hrule}%
    \hbox{\kern.1ex\box1\kern.1ex}
  }
\newcommand{\bisim}{\mathrel{\sbisim\hspace*{1pt}}}

\newcommand{\sfunbisimos}{%
    \setbox0=\hbox{\kern-.1ex{$\rightarrow$}\kern-.1ex}
    \setbox1=\vbox{\hbox{\raise .1ex \box0}\hrule}%
    {\pap{\hbox{\kern.05ex\box1\kern.1ex}}{\hspace*{0.5pt}\subosr}}
  }

\newcommand{\sconvfunbisimos}{%
    \setbox0=\hbox{\kern-.1ex{$\leftarrow$}\kern-.1ex}
    \setbox1=\vbox{\hbox{\raise .1ex \box0}\hrule}%
    {\pap{\hbox{\kern.05ex\box1\kern.1ex}}{\hspace*{0.5pt}\subosr}}
  }

\newcommand{\sbisimos}{%
    \setbox0=\hbox{\kern-.1ex{$\leftrightarrow$}\kern-.1ex}
    \setbox1=\vbox{\hbox{\raise .1ex \box0}\hrule}%
    \ensuremath{\pap{\mathrel{\hbox{\kern.1ex\box1\kern.1ex}}}{\hspace*{0.5pt}\subosr}}
  }

\newcommand{\sfunbisimosvia}[1]{%
    \setbox0=\hbox{\kern-.1ex{$\rightarrow$}\kern-.1ex}
    \setbox1=\vbox{\hbox{\raise .1ex \box0}\hrule}%
    {\bpap{\hbox{\kern.05ex\box1\kern.1ex}}{#1}{\hspace*{0.5pt}\subosr}}
  }

\newcommand{\sconvfunbisimosvia}[1]{%
    \setbox0=\hbox{\kern-.1ex{$\leftarrow$}\kern-.1ex}
    \setbox1=\vbox{\hbox{\raise .1ex \box0}\hrule}%
    {\bpap{\hbox{\kern.05ex\box1\kern.1ex}}{#1}{\hspace*{0.5pt}\subosr}}
  }

\newcommand{\avar}{x}

\newcommand{\sslabs}{\lambda}
\newcommand{\slabs}[1]{\sslabs{#1}}
\newcommand{\labs}[2]{\slabs{#1}.\,{#2}}

\newcommand{\slapp}{\hspace*{1.5pt}}
\newcommand{\lapp}[2]{{#1}\slapp{#2}} 

\newcommand{\avert}{v}

\definecolor{azure}{rgb}{0.94,1.00,1.00}
\definecolor{brown}{rgb}{.75,.25,.25}
\definecolor{cyan}{rgb}{0.25,0.88,0.82}
\definecolor{chocolate}{rgb}{0.82,0.41,0.12}
\definecolor{darkcyan}{rgb}{0.5,0,1}
\definecolor{darkgreen}{rgb}{0,0.39,0}
\definecolor{darkmagenta}{rgb}{0.5,0,0.5}
\definecolor{darkgoldenrod}{RGB}{184,134,11}
\definecolor{firebrick}{RGB}{175,25,25}
\definecolor{forestgreen}{rgb}{0.13,0.55,0.13}
\definecolor{goldenrod}{RGB}{218,165,32}
\definecolor{grey}{RGB}{72,72,72}
\definecolor{lightcyan}{rgb}{0.88,1.00,1.00}
\definecolor{lightpink}{rgb}{1.00,0.71,0.76}
\definecolor{myyellow}{RGB}{235,235,0}
\definecolor{lightyellow}{rgb}{1.00,1.00,0.88}
\definecolor{lightgoldenrod}{rgb}{0.83,0.97,0.51}
\definecolor{lightgoldenrodyellow}{rgb}{0.98,0.98,0.82}
\definecolor{lightskyblue}{rgb}{0.53,0.81,0.98}
\definecolor{moccasin}{rgb}{1.00,0.89,0.71}
\definecolor{magenta}{rgb}{1,0,1}
\definecolor{navyblue}{rgb}{0,0,0.5}
\definecolor{orange}{rgb}{1.0,0.65,0.0}
\definecolor{orangered}{rgb}{1.0,0.27,0.0}
\definecolor{palegreen}{rgb}{0.60,0.98,0.60}
\definecolor{powderblue}{rgb}{0.69,0.88,0.90}
\definecolor{purple}{rgb}{1,0.5,1}
\definecolor{royalblue}{RGB}{65,105,225}
\definecolor{mediumblue}{RGB}{0,0,205}
\definecolor{cornflowerblue}{RGB}{100,149,237}
\definecolor{springgreen}{rgb}{0.0,1.0,0.5}
\definecolor{turquoise}{rgb}{0.25,0.88,0.82}
\definecolor{snow}{rgb}{1.00,0.98,0.98}
\definecolor{tan}{rgb}{0.82,0.71,0.55}
\definecolor{red}{rgb}{1,0,0}
\definecolor{violetred}{RGB}{208,32,144}

\newcommand{\colorin}[1]{\textcolor{#1}}

\newcommand{\forestgreen}{\colorin{forestgreen}}

\newcommand{\mediumblue}{\colorin{mediumblue}}

\newcommand{\lambdacalculus}{$\lambda$\nb-cal\-cu\-lus}

\newcommand{\lambdaterm}{$\lambda$\nb-term}
\newcommand{\lambdaterms}{\lambdaterm{s}}

\newcommand{\loopentry}{loop-en\-try}

\newcommand{\nonequational}{non-equa\-ti\-onal}

\newcommand{\onetransition}{$\sone$\nb-tran\-si\-tion}
\newcommand{\onetransitions}{\onetransition{s}}

\newcommand{\structureconstrained}{struc\-ture-con\-strained}

%% file: figs/pipeline-1.tex
  \begin{tikzpicture}
    \pgfdeclarelayer{background}
    \pgfdeclarelayer{outerscope}
    \pgfdeclarelayer{innerscope}
    \pgfdeclarelayer{graph}
    \pgfdeclarelayer{boundingbox}
    \pgfdeclarelayer{caption}
    \pgfsetlayers{boundingbox,caption,background,outerscope,innerscope,graph}
    \input{figs/L0-syntax-tree-0.tex}
    \input{figs/L0-syntax-tree-2.tex}
    \input{figs/L0-syntax-tree-5.tex}
    \input{figs/L0-lhotg-1.tex}
    \input{figs/L0-lhotg-3a.tex}
    \input{figs/L0-lhotg-4a.tex}
  \end{tikzpicture}

%% file: figs/pipeline-2.tex
  \begin{tikzpicture}
    \pgfdeclarelayer{background}
    \pgfdeclarelayer{outerscope}
    \pgfdeclarelayer{innerscope}
    \pgfdeclarelayer{graph}
    \pgfdeclarelayer{boundingbox}
    \pgfdeclarelayer{caption}
    \pgfsetlayers{boundingbox,caption,background,outerscope,innerscope,graph}
    \input{figs/L0-lhotg-4.tex}      
    \input{figs/L0-lhotg-3.tex}
    \input{figs/L0-ltg-1.tex}
    \input{figs/L0-ltg-2.tex}
    \input{figs/L0-lnfa-2.tex}
    \input{figs/L0-ldfa.tex}
  \end{tikzpicture}

%% file: figs/expressible.tex
\begin{tikzpicture}

\matrix[anchor=north,row sep=1cm,%
        every node/.style={draw,thick,circle,minimum width=2.5pt,fill,inner sep=0pt,outer sep=2pt},%
        ampersand replacement=\&] 
        at (0,0) {
  \node(C1-0){};
  \\
  \node(C1-1){};
  \\
  \node(C1-2){};
  \\
  \node(C1-3){};
  \\      
};

\matrix[anchor=north,row sep=1cm,%
        every node/.style={draw,thick,circle,minimum width=2.5pt,fill,inner sep=0pt,outer sep=2pt},%
        ampersand replacement=\&] 
        at (4.25,-0.7) {
  \node(2E-0){};
  \\
  \node(2E-1){};
  \\
  \node(2E-2){};
  \\      
}; 
\matrix[anchor=north,row sep=1cm,%
        every node/.style={draw,thick,circle,minimum width=2.5pt,fill,inner sep=0pt,outer sep=2pt},%
        ampersand replacement=\&] 
        at (8.5,0) {
  \node(C2-0){};
  \\
  \node(C2-1){};
  \\
  \node(C2-2){};
  \\
  \node(C2-3){};
  \\      
}; 

\draw[<-,thick,chocolate,>=latex](C1-0) -- ++ (90:0.4cm);
\draw[->] (C1-0) to node[right]{$\!{\scriptstyle a}$} (C1-1);
\draw[->] (C1-1) to node[right]{$\!{\scriptstyle a}$} (C1-2);
\draw[->] (C1-2) to node[right]{$\!{\scriptstyle b}$} (C1-3);
\draw[shorten >=3pt,shorten <=3pt] ([shift=(270:1.17cm)]C1-2) arc[radius=1.17cm,start angle=270,end angle=90];
\path (C1-2) ++ (-1.3cm,0cm) node{${\scriptstyle a}$};
\draw[->,shorten >=3pt,shorten <=3pt] ([shift=(270:1.13cm)]C1-1) arc[radius=0.565cm,start angle=270,end angle=90];
\path ([shift=(90:0.575cm)]C1-2) ++ (-0.69cm,0cm) node{${\scriptstyle b}$};
\path (2E-1) ++ (-4.5cm,-2.6cm) node{$\procsem{a\cdot{(a\cdot(b + b\cdot a))^*} \cdot 0} \in \fap{\mathit{im}}{\procsem{\cdot}}$};
\path (2E-1) ++ (-4.25cm,-3.2cm)  node{\procsemexpressible};
\draw[<-,thick,chocolate,>=latex](2E-0) -- ++ (90:0.4cm);
\draw[->] (2E-0) to node[right]{$\!{\scriptstyle a}$} (2E-1);
\draw[->] (2E-1) to node[right]{$\!{\scriptstyle a}$} (2E-2);
\draw[->,shorten >=3pt,shorten <=3pt] ([shift=(270:1.17cm)]2E-1) arc[radius=1.17cm,start angle=-90,end angle=90];
\path (2E-1) ++ (1.3cm,0cm) node{${\scriptstyle b}$};
\draw[->,out=180,in=180,distance=1.05cm] (2E-2) to node[left]{$\scriptstyle b$} (2E-1);
%
\draw[<-,thick,chocolate,>=latex](C2-0) -- ++ (90:0.4cm);
\draw[->] (C2-0) to node[right]{$\!{\scriptstyle a}$} (C2-1);
\draw[->] (C2-1) to node[right]{$\!{\scriptstyle a}$} (C2-2);
\draw[->] (C2-2) to node[right]{$\!{\scriptstyle b}$} (C2-3);
\draw[->,shorten >=3pt,shorten <=3pt] ([shift=(0:0pt)]C2-2) arc[radius=1.17cm,start angle=-90,end angle=90];
\path (C2-1) ++ (1.3cm,0cm) node{${\scriptstyle b}$};
\draw[->,shorten >=3pt,shorten <=3pt] ([shift=(270:1.16cm)]C2-2) arc[radius=0.58cm,start angle=-90,end angle=75];
\path ([shift=(270:0.575cm)]C2-2) ++ (0.72cm,0cm) node{${\scriptstyle a}$};
\path (2E-1) ++ (4.5cm,-2.6cm) node{$\procsem{(a \cdot a \cdot {(b \cdot a)^*} \cdot b)^* \cdot 0} \in \fap{\mathit{im}}{\procsem{\cdot}}$};
\path (2E-1) ++ (4.25cm,-3.2cm) node{\procsemexpressible};
%
%
\draw[color=magenta,densely dashed] (C1-0) to (2E-0);
\draw[color=magenta,densely dashed] (C1-1) to (2E-1);
\draw[color=magenta,densely dashed] (C1-2) to (2E-2);
\draw[color=magenta,densely dashed] (C1-3) to (2E-0);
%
%
%
\draw[color=magenta,densely dashed] (C2-0) to (2E-0);
\draw[color=magenta,densely dashed] (C2-1) to (2E-1);
\draw[color=magenta,densely dashed] (C2-2) to (2E-2);
\draw[color=magenta,densely dashed] (C2-3) to (2E-1);
%
\path (2E-1) ++ (0cm,-2.6cm) node{$\notin \fap{\mathit{im}}{\procsem{\cdot}}$};
\path (2E-1) ++ (0cm,-3.2cm) node{\procsemexpressible};
\path (2E-1) ++ (0cm,-3.6cm) node{modulo $\sbisim$};

\end{tikzpicture}

%% file: figs/not-expressible-mod-bisim.tex
\begin{tikzpicture}


\matrix[anchor=north,row sep=0.8cm,every node/.style={draw,thick,circle,minimum width=2.5pt,fill,inner sep=0pt,outer sep=2pt}] at (5.5,0) {
  \node(C-11-0){}; 
  \\
  \node[opacity=0](helper-11){};
  \\
  \node(C-11-1){};
  \\
};

\draw[<-,thick,chocolate,>=latex,shorten <= 2pt](C-11-0) -- ++ (90:0.4cm ++ 2.5pt);

\draw[thick] (C-11-0) circle (0.12cm);
\draw[->,bend right,shorten <= 2.5pt,shorten >= 2.5pt] (C-11-0) to node[left]{${\scriptstyle a}\!$} (C-11-1);
                                                                
\draw[thick] (C-11-1) circle (0.12cm); 
\draw[->,bend right,shorten <= 2.5pt,shorten >= 2.5pt] (C-11-1) to node[right]{$\!{\scriptstyle b}$} (C-11-0);



\matrix[anchor=north,row sep=0.8cm,column sep=0.924cm,ampersand replacement=\&,
        every node/.style={draw,very thick,circle,minimum width=2.5pt,fill,inner sep=0pt,outer sep=2pt}] at (11,0) {
                   \& \node(C-21-0){};               \&
  \\
                   \& \node[opacity=0](helper-2-C21){}; \&                  
  \\
  \node(C-21-1){}; \& \node[opacity=0](helper-2-C21){}; \&     \node(C-21-2){};
  \\
};
\draw[<-,thick,chocolate,>=latex](C-21-0) -- ++ (90:0.4cm);

\draw[->,bend right,distance=0.6cm] (C-21-0) to node[left,xshift=0.2ex]{${\scriptstyle a_1}$} (C-21-1); 
\draw[->,bend left,distance=0.6cm,] (C-21-0) to node[right,xshift=-0.2ex]{${\scriptstyle a_2}$} (C-21-2);

\draw[->,bend right,distance=0.6cm] (C-21-1) to node[left,xshift=0.3ex,pos=0.55]{${\scriptstyle b_1}$} (C-21-0); 
\draw[->,bend left,distance=0.6cm]  (C-21-1) to node[above,yshift=-0.325ex]{${\scriptstyle b_2}$} (C-21-2); 

\draw[->,bend left,distance=0.6cm,shorten <= 9pt] (C-21-2) to node[right,xshift=-0.325ex,pos=0.55]{${\scriptstyle c_1}$} (C-21-0); 
\draw[->,bend left,distance=0.6cm]  (C-21-2) to node[above,yshift=-0.2ex]{${\scriptstyle c_2}$} (C-21-1);



\end{tikzpicture}

%% file: figs/LEE-extraction-2-LEE.tex
\begin{tikzpicture}
\matrix[anchor=center,row sep=1cm,every node/.style={draw,thick,circle,minimum width=2.5pt,fill,inner sep=0pt,outer sep=2pt}] at (0,0) {
  \node(C-0){}; 
  \\
  \node(C-1){};
  \\
  \node(C-2){};
  \\
};
\draw[<-,thick,chocolate,>=latex](C-0) -- ++ (90:0.4cm);
\draw[->] (C-0) to node[right]{$\!{\scriptstyle a}$} (C-1);
%
\draw[->,forestgreen] (C-1) to node[right]{$\!{\scriptstyle a}$} (C-2);
%
\draw[->,shorten >=3pt,shorten <=3pt] ([shift=(270:1.13cm)]C-1) arc[radius=1.13cm,start angle=-90,end angle=90];
%
\draw[->,very thick,color=forestgreen,shorten >=3pt,shorten <=3pt] ([shift=(270:1.13cm)]C-1) arc[radius=0.57cm,start angle=270,end angle=90];
\path (C-1) ++ (1.25cm,0cm) node{${\scriptstyle b}$};
\path ([shift=(270:0.575cm)]C-1) ++ (-0.72cm,0cm) node{$\forestgreen{\scriptstyle b}$};
%
%

%
%
%
%
\matrix[anchor=center,row sep=1cm,every node/.style={draw,thick,circle,minimum width=2.5pt,fill,inner sep=0pt,outer sep=2pt}] at (4.75,0) {
  \node(C'-0){}; 
  \\
  \node(C'-1){};
  \\
  \node(C'-2){};
  \\
};
\draw[<-,thick,chocolate,>=latex](C'-0) -- ++ (90:0.4cm);
\draw[->,very thick,forestgreen] (C'-0) to node[right]{$\!{\scriptstyle a}$} (C'-1);
%
\draw[->] (C'-1) to node[right]{$\!{\scriptstyle a}$} (C'-2);
\draw[->,color=forestgreen] (C'-1) to node[right]{$\!{\scriptstyle a}$} (C'-2);
\draw[->,shorten >=3pt,shorten <=3pt] ([shift=(270:1.13cm)]C'-1) arc[radius=1.13cm,start angle=-90,end angle=90];
\draw[->,color=forestgreen,shorten >=3pt,shorten <=3pt] ([shift=(270:1.13cm)]C'-1) arc[radius=1.13cm,start angle=-90,end angle=90];
\path (C'-1) ++ (1.25cm,0cm) node[forestgreen]{${\scriptstyle b}$};
         
\draw (C-1) edge[shorten <=8.5ex,shorten >= 2ex,-implies,double equal sign distance] 
              node[above,pos=0.6]{{\scriptsize eliminate loop entry}} 
              node[below,pos=0.6]{{\scriptsize eliminate loop}}
      (C'-1);  
%
%
%
%
\matrix[anchor=center,row sep=1cm,every node/.style={draw,thick,circle,minimum width=2.5pt,fill,inner sep=0pt,outer sep=2pt}] at (9.5,0) {
  \node(C''-0){}; 
  \\
  \node(C''-1){};
  \\
  \node(C''-2){};
  \\
};
\draw[<-,thick,chocolate,>=latex](C''-0) -- ++ (90:0.4cm);
\draw[->] (C''-1) to node[right]{$\!{\scriptstyle a}$} (C''-2);
\draw[->,shorten >=3pt,shorten <=3pt] ([shift=(270:1.13cm)]C''-1) arc[radius=1.13cm,start angle=-90,end angle=90];
\path (C''-1) ++ (1.25cm,0cm) node{${\scriptstyle b}$};
\draw (C'-1) edge[shorten <=8.5ex,shorten >= 2ex,-implies,double equal sign distance] 
              node[above,pos=0.6]{{\scriptsize eliminate loop entry}}
      (C''-1);  
\matrix[anchor=center,row sep=1cm,every node/.style={draw,thick,circle,minimum width=2.5pt,fill,inner sep=0pt,outer sep=2pt}] at (13.5,0) {
  \node(C'''-0){};
  \\
  \node[draw=none,fill=none](C'''-1){};
  \\
  \node[draw=none,fill=none](C'''-2){};
  \\
};
\draw[<-,thick,chocolate,>=latex](C'''-0) -- ++ (90:0.4cm);
%
%
%
\draw (C''-1) edge[shorten <=8.5ex,shorten >= 0ex,-implies,double equal sign distance] 
              node[above,pos=0.675]{{\scriptsize garbage collection}}
      (C'''-1);
      
\draw (C'-1) edge[shorten <=8.5ex,shorten >= 2ex,-implies,double equal sign distance,bend right,distance=4cm,in=215] 
              node[below,pos=0.6,yshift=-1ex]{{\scriptsize eliminate loop}}
      (C'''-1);

%
%
%
%
%
%
%
%
%
%
%
%
%
%
\matrix[anchor=center,row sep=1cm,every node/.style={draw=none}] at (0,-4) {
  \node[draw,thick,circle,minimum width=2.5pt,fill,inner sep=0pt,outer sep=2pt](C-synth-0){}; 
  \\
  \node(C-synth-1){};
  \\
  \node(C-synth-2){};
  \\
};
\path (C-synth-0) ++ (0.55cm,0cm) node[draw,very thick,circle,minimum width=2.5pt,fill,inner sep=0pt,outer sep=2pt](C-synth-0'){};
\path (C-synth-1) ++ (0.55cm,0cm) node[draw,very thick,circle,minimum width=2.5pt,fill,inner sep=0pt,outer sep=2pt](C-synth-1'){};
\path (C-synth-2) ++ (0.55cm,0cm) node[draw,very thick,circle,minimum width=2.5pt,fill,inner sep=0pt,outer sep=2pt](C-synth-2'){};
\draw[<-,thick,chocolate,>=latex](C-synth-0) -- ++ (90:0.4cm);
\draw[->,thick,forestgreen] (C-synth-0') to node[right]{$\!{\scriptstyle a}$} (C-synth-1');
\draw[->] (C-synth-1') to node[right]{$\!{\scriptstyle a}$} (C-synth-2');
\draw[densely dotted,thick,bend left,distance=0.15cm] (C-synth-0) to (C-synth-0');
\draw[->,shorten >=3pt,shorten <=3pt] ([shift=(270:1.23cm)]C-synth-1') arc[radius=1.21cm,start angle=-90,end angle=90];
\path (C-synth-1') ++ (1cm,0cm) node{${\scriptstyle b}$}; 
\path (C-synth-1') ++ (1.9cm,0cm) node[draw,thick,forestgreen,circle,minimum width=2.5pt,fill,inner sep=0pt,outer sep=2pt](C-synth-1''){};
\path (C-synth-2') ++ (1.9cm,0cm) node[draw,thick,forestgreen,circle,minimum width=2.5pt,fill,inner sep=0pt,outer sep=2pt](C-synth-2''){};
\draw[->,thick,forestgreen,shorten >=3pt,shorten <=3pt] ([shift=(270:1.21cm)]C-synth-1'') arc[radius=0.6cm,start angle=270,end angle=90];
\path ([shift=(270:0.575cm)]C-synth-1'') ++ (-0.45cm,0cm) node[forestgreen]{${\scriptstyle b}$};
\draw[->] (C-synth-1'') to node[right]{$\!{\scriptstyle a}$} (C-synth-2'');
\draw[densely dotted,thick,bend right,distance=0.5cm] (C-synth-2') to (C-synth-2'');
%
%

\matrix[anchor=center,row sep=1cm,every node/.style={draw=none}] at (4.75,-4) {
  \node[draw,thick,circle,minimum width=2.5pt,fill,inner sep=0pt,outer sep=2pt](C'-synth-0){}; 
  \\
  \node(C'-synth-1){};
  \\
  \node(C'-synth-2){};
  \\
};
\path (C'-synth-0) ++ (0.55cm,0cm) node[draw,thick,circle,minimum width=2.5pt,fill,inner sep=0pt,outer sep=2pt](C'-synth-0'){};
\path (C'-synth-1) ++ (0.55cm,0cm) node[draw,thick,circle,minimum width=2.5pt,fill,inner sep=0pt,outer sep=2pt](C'-synth-1'){};
\path (C'-synth-2) ++ (0.55cm,0cm) node[draw,thick,circle,minimum width=2.5pt,fill,inner sep=0pt,outer sep=2pt](C'-synth-2'){};
\draw[<-,thick,chocolate,>=latex](C'-synth-0) -- ++ (90:0.4cm);
\draw[->,thick,forestgreen] (C'-synth-0') to node[right]{$\!{\scriptstyle a}$} (C'-synth-1');
\draw[->] (C'-synth-1') to node[right]{$\!{\scriptstyle a}$} (C'-synth-2');
\draw[densely dotted,thick,bend left,distance=0.15cm] (C'-synth-0) to (C'-synth-0');
\draw[->,shorten >=3pt,shorten <=3pt] ([shift=(270:1.22cm)]C'-synth-1') arc[radius=1.20cm,start angle=-90,end angle=90];
\path (C'-synth-1') ++ (1.3cm,0cm) node{${\scriptstyle b}$};

\draw (C-synth-1) edge[shorten <=16ex,shorten >=1ex,implies-,double equal sign distance] 
                   node[below,pos=0.8]{{\scriptsize add loop}}
      (C'-synth-1);

%
%
%
%
%
%
\matrix[anchor=center,row sep=1cm] at (13.5,-4) {
  \node[draw,thick,circle,minimum width=2.5pt,fill,inner sep=0pt,outer sep=2pt](C''-synth-0){};
  \\
  \node(C''-synth-1){};
  \\
  \node(C''-synth-2){};
  \\ 
};
\draw[<-,thick,chocolate,>=latex](C''-synth-0) -- ++ (90:0.4cm);

\draw (C'-synth-1) edge[shorten <=14ex,shorten >= 4ex,implies-,double equal sign distance] 
                   node[below,pos=0.59]{{\scriptsize add loop}}
      (C''-synth-1);
\end{tikzpicture}

%% file: figs/LEE-extraction-2-result.tex
\begin{tikzpicture}
\matrix[anchor=north,row sep=1cm,every node/.style={draw,thick,circle,minimum width=2.5pt,fill,inner sep=0pt,outer sep=2pt}] at (0,0) {
  \node(C-0){}; 
  \\
  \node(C-1){};
  \\
  \node(C-2){};
  \\
};
\draw[<-,thick,color=chocolate,>=latex](C-0) -- ++ (90:0.4cm);
%
\draw[->,thick,color=forestgreen] (C-0) to node[right,color=forestgreen]{$\!{\scriptstyle a}$} node[left,color=forestgreen]{${\scriptstyle [2]\hspace*{-0.5ex}}$} (C-1);
\draw[->] (C-1) to node[right]{$\!{\scriptstyle a}$} (C-2);
\draw[->,shorten >=3pt,shorten <=3pt] ([shift=(270:1.13cm)]C-1) arc[radius=1.13cm,start angle=-90,end angle=90];
\draw[->,shorten >=3pt,shorten <=3pt] ([shift=(270:1.13cm)]C-1) arc[radius=0.565cm,start angle=270,end angle=90];
\draw[->,thick,forestgreen,shorten >=3pt,shorten <=3pt] ([shift=(270:1.13cm)]C-1) arc[radius=0.565cm,start angle=270,end angle=90];
\path (C-1) ++ (1.25cm,0cm) node{${\scriptstyle b}$};
\path ([shift=(270:0.575cm)]C-1) ++ (-0.42cm,0cm) node[forestgreen]{${\scriptstyle b}$};
\path ([shift=(270:0.575cm)]C-1) ++ (-0.8cm,0cm) node[color=forestgreen]{${\scriptstyle [1]}$};

\matrix[anchor=north,row sep=1cm,every node/.style={draw=none}] at (3.5,0) {
  \node[draw,thick,circle,minimum width=2.5pt,fill,inner sep=0pt,outer sep=2pt](C-synth-0){}; 
  \\
  \node(C-synth-1){};
  \\
  \node(C-synth-2){};
  \\
};
\path (C-synth-0) ++ (0.55cm,0cm) node[draw,thick,circle,minimum width=2.5pt,fill,inner sep=0pt,outer sep=2pt](C-synth-0'){};
\path (C-synth-1) ++ (0.55cm,0cm) node[draw,thick,circle,minimum width=2.5pt,fill,inner sep=0pt,outer sep=2pt](C-synth-1'){};
\path (C-synth-2) ++ (0.55cm,0cm) node[draw,thick,circle,minimum width=2.5pt,fill,inner sep=0pt,outer sep=2pt](C-synth-2'){};
\draw[<-,thick,color=chocolate,>=latex](C-synth-0) -- ++ (90:0.4cm);
\draw[->,thick,forestgreen] (C-synth-0') to node[right]{$\!{\scriptstyle a}$} (C-synth-1');
\draw[->] (C-synth-1') to node[right]{$\!{\scriptstyle a}$} (C-synth-2');
\draw[densely dotted,thick,bend left,distance=0.15cm] (C-synth-0) to (C-synth-0');
\draw[->,shorten >=3pt,shorten <=3pt] ([shift=(270:1.23cm)]C-synth-1') arc[radius=1.21cm,start angle=-90,end angle=90];
\path (C-synth-1') ++ (1cm,0cm) node{${\scriptstyle b}$}; 
\path (C-synth-1') ++ (1.9cm,0cm) node[draw,thick,circle,minimum width=2.5pt,fill,inner sep=0pt,outer sep=2pt](C-synth-1''){};
\path (C-synth-2') ++ (1.9cm,0cm) node[draw,thick,circle,minimum width=2.5pt,fill,inner sep=0pt,outer sep=2pt](C-synth-2''){};
\draw[->,thick,forestgreen,shorten >=3pt,shorten <=3pt] ([shift=(270:1.21cm)]C-synth-1'') arc[radius=0.6cm,start angle=270,end angle=90];
\path ([shift=(270:0.575cm)]C-synth-1'') ++ (-0.45cm,0cm) node[forestgreen]{${\scriptstyle b}$};
\draw[->] (C-synth-1'') to node[right]{$\!{\scriptstyle a}$} (C-synth-2'');
\draw[densely dotted,thick,bend right,distance=0.5cm] (C-synth-2') to (C-synth-2'');
\matrix[anchor=north,row sep=1cm,%
        every node/.style={draw,thick,circle,minimum width=2.5pt,fill,inner sep=0pt,outer sep=2pt},%
        ampersand replacement=\&] 
        at (9.25,0) {
  \node(C2-0){};
  \\
  \node(C2-1){};
  \\
  \node(C2-2){};
  \\
  \node(C2-3){};
  \\      
};    
\draw[<-,thick,color=chocolate,>=latex](C2-0) -- ++ (90:0.4cm);
\draw[->,thick,forestgreen] (C2-0) to node[right]{$\!{\scriptstyle a}$} (C2-1);
\draw[->] (C2-1) to node[right]{$\!{\scriptstyle a}$} (C2-2);
\draw[->,thick,forestgreen] (C2-2) to node[right]{$\!{\scriptstyle b}$} (C2-3);
\draw[->,shorten >=3pt,shorten <=3pt] ([shift=(0:0pt)]C2-2) arc[radius=1.17cm,start angle=-90,end angle=90];
\path (C2-1) ++ (1.3cm,0cm) node{${\scriptstyle b}$};
\draw[->,shorten >=3pt,shorten <=3pt] ([shift=(270:1.16cm)]C2-2) arc[radius=0.58cm,start angle=-90,end angle=75];
\path ([shift=(270:0.575cm)]C2-2) ++ (0.74cm,0cm) node{${\scriptstyle a}$};
\path (C-2) ++ (0cm,-2cm) node{\scalebox{1}{\mediumblue{LEE-witness}}};

\path (C-synth-2) ++ (1.2cm,-1.675cm) node{\parbox{\widthof{LEE-witness}}
                                                 {\hspace*{\fill}\mediumblue{structured}\hspace*{\fill}\mbox{}
                                                  \\[-0.25ex]
                                                  \mediumblue{LEE-witness}}}; 

\path (C2-2) ++ (0.2cm,-2cm) node{\scalebox{1}{$\procsem{(a a (b a)^* b)^* 0}$}};
\end{tikzpicture}